\documentclass[aps,preprint,epsfig,rotate]{revtex4}
\begin{document}
%\begin{doublespace}
\title{On the Galilean transformation of the few-electron wave functions.}

 \author{Alexei M. Frolov}
 \email[E--mail address: ]{afrolov@uwo.ca}

\affiliation{Department of Chemistry\\
 University of Western Ontario, London, Ontario N6H 5B7, Canada}

\date{\today}

\begin{abstract}

The Galilean transformations of the few-electron atomic wave functions are 
considered. We discuss the few-electron wave functions constructed in the model of 
independent electrons as well as the truly correlated (or highly accurate) wave 
functions. Results of our analysis are applied to determine the probability of
formation of the negatively charged tritium/protium ions during the nuclear 
$(n,{}^{3}$He$;t,p)-$reaction of the helium-3 atoms with thermal/slow neutrons. 

\end{abstract}

\maketitle
\newpage

The general formulas for the Galilean transformation of the non-relativistic, 
single-particle wave function $\Psi({\bf r}, t)$ are well known since the middle of 
1920's \cite{Ehre}. A complete discussion of this problem can be found, e.g., in
\cite{LLQ} and \cite{Kaemp}. If $\Psi({\bf r}, t)$ is such a wave function 
written in the reference frames which are at rest, then the corresponding wave 
function $\Psi^{\prime}({\bf r}^{\prime}, t)$ in the moving frames takes the form
\begin{eqnarray}
 \Psi^{\prime}({\bf r} - {\bf V} t, t) = \Psi({\bf r}, t)
 exp\Bigl[ \frac{\imath m}{\hbar} (-{\bf V} \cdot {\bf r} + \frac12 V^2 t) \Bigr]
 \label{eq1}
\end{eqnarray}
or
\begin{eqnarray}
 \Psi^{\prime}({\bf r}, t) = \Psi({\bf r} + {\bf V} t, t)
 exp\Bigl[ \frac{\imath m}{\hbar} (-{\bf V} \cdot {\bf r} - \frac12 V^2 t) \Bigr]
 \label{eq2}
\end{eqnarray}
From Eq.(\ref{eq2}) one easily finds the following transformation formulas for the 
non-relativistic wave function of an arbitrary few-particle system. To avoid analysis
of very general quantum systems in this study we restrict ourselves to the consideration 
of few-electron atomic systems only. By an `atomic system' we mean the bound few-electron 
atom with one very heavy center which also has a positive electric charge $Qe$. The forces 
of electric attraction between nucleus and electrons bind this system together. The 
competing forces of electric repulsion between electrons decrease the final value of 
binding energy, but they are relatively small and cannot destroy the whole atom, or
produce electron ionization in it. 

The general formulas for the Galilean transformations of the actual few-electron (atomic) 
wave function follow from formulas, Eqs.(\ref{eq1}) - (\ref{eq2}). For such systems one
finds a number of advantages to write all formulas in atomic units, where $\hbar = 1, e 
= 1, m_e = 1$. In this units for the $N-$electron atomic system we have
\begin{eqnarray}
 \Psi^{\prime}({\bf r}_1, \ldots, {\bf r}_N, t) = \Psi({\bf r}_1 + {\bf V} t, \ldots,
 {\bf r}_N + {\bf V} t, t) exp\Bigl[ \imath (-{\bf V} \cdot \sum^{N}_{i=1}{\bf r}_i - N 
 \frac12 V^2 t) \Bigr] \label{eq3}
\end{eqnarray}
where all electrons are assumed to be independent, or non-correlated. By approximating the 
actual $N-$electron wave functions by the trial functions constructed in the model of 
independent electrons one can show that the formula, Eq.(\ref{eq3}), is also correct in the 
general case, i.e. when all electron-electron correlations are included. Briefly, we can say
that the Galilean transformation of the actual (i.e. truly correlated) wave function is 
represented by the same formula, Eq.(\ref{eq3}), where the phase factor does not depend (and 
cannot depend) upon any of the electron-electron coordinates. 

In reality, we need the formulas for the Galilean transformations of the $N-$electron 
wave functions in the limit $t \rightarrow 0$. This limit corresponds to the sudden 
approximation \cite{Mig1}, \cite{Mig2} for few-electron atomic systems. For instance, all
processes, decays and reactions in atomic nuclei proceed significantly faster than usual 
electron transitions in atoms. Therefore, the sudden approximation can be applied  to
determine the probabilities of the electron-electron transitions during nuclear reactions 
in atoms. In many cases it is important to know the probabilities to form various `final 
atomic states' after some fast nuclear process in the atomic nucleus. The `incident' atomic
state is usualy known. In many cases the newly created nuclei (or `nuclear fragments') are 
rapidly moving after the nuclear reaction and/or decay. In such cases one finds numerous 
advantages to determine the final state probabilities by using the moving frames with the 
origin located at the central atomic nucleus. In these frames if the nucleus begins to move,
then all atomic electrons suddenly take the speed $-{\bf V}_n$, where the subscript $n$ means 
the nucleus. Now, the formula for the sudden (Galilean) transformation of the non-relativistic 
wave function of a $N$-electron atomic system takes the form 
\begin{eqnarray}
 \Psi^{\prime}({\bf r}_1, \ldots, {\bf r}_N) = 
 exp(\imath {\bf V}_n \cdot {\bf r}_1 + \imath {\bf V}_n \cdot {\bf r}_2 + \ldots
 + \imath {\bf V}_n \cdot {\bf r}_N) \Psi({\bf r}_1, \ldots, {\bf r}_N) \label{eq4}
\end{eqnarray}
This formula is written in the form which can directly be used for an arbitrary 
$N-$electron atom with independent and/or quasi-independent electrons. In such cases the 
wave function depends upon the $N$ electron-nuclear $r_i = r_{in}$ coordinates (scalars) 
only. However, all actual, few-electron wave functions are truly correlated, i.e. they 
explicitly depend upon both the electron-nuclear $r_i = r_{in}$ and electron-electron $r_{ij}$ 
coordinates, which are also called the interparticle coordinates (or Hylleraas coordinates). 
Formulas for the Galilean transformations of the electron-electron coordinates $r_{ij}$ can be 
obtained from their definitions. Indeed, according to the definition of $r_{ij}$ we can write 
the following identities $r_{ij} = \mid {\bf r}_i - {\bf r}_j \mid = \mid {\bf r}_i - {\bf r}_n 
- ({\bf r}_j - {\bf r}_n) \mid = \mid {\bf r}_i - ({\bf r}_n - {\bf V} \delta t) - [{\bf r}_j 
- ({\bf r}_n - {\bf V} \delta t)] \mid = r_{ij}$, where $\delta t$ is infinitely small. This 
means that the electron-electron coordinates $r_{ij}$ does not depend upon ${\bf V}$, i.e. it 
does not change during the sudden motion of the nucleus. It follows from here that the sudden 
Galilean transformations of the truly correlated wave function are also described by the same 
formula, Eq.(\ref{eq4}).
  
The formula, Eq.(\ref{eq4}), can be applied to determine the probability of formation of the 
tritium ${}^3$H$^{-}$ ion during the reaction of the ${}^3$He nuclei with slow/thermal neutrons
\cite{Kik}
\begin{eqnarray}
 {}^{3}{\rm He} + n = {}^3{\rm H} + {}^1{\rm H} + 0.764 \; \; \; MeV \label{en1}
\end{eqnarray}
in the two-electron helium-3 atom. In the reaction Eq.(\ref{en1}) the notations ${}^3$H and 
${}^1{\rm H}$ stand for the tritium nucleus (or $t$ nucleus) and protium (or $p$ nucleus).  
The reaction, Eq.(\ref{en1}), is of great interest for the burning of the high-dense ($\rho \ge$ 
100 $g \cdot cm^{-3}$) deuterium plasmas \cite{Gon}, \cite{Fro98}. The reaction, Eq.(\ref{en1}), 
in the two-electron ${}^3$He atom and one-electron ${}^3$He$^{+}$ ion was considered in our earlier 
studies \cite{Fro1}, \cite{Fro2}. The cross-section $\sigma$ of this nuclear reaction for thermal 
neutrons with $E_n \approx 0$ is very large $\sigma_{max} \approx 5330 \cdot 10^{-24}$ $cm^2$ (or 
5330 $barn$) \cite{WW}. The velocities of the two nuclear fragments formed in the reaction, 
Eq.(\ref{en1}), with thermal neutrons are $v_t \approx$ 1.59632 $a.u.$ and $v_{p} \approx$ 4.78797 
$a.u.$ for the tritium and protium nuclei, respectively. In this study all particle velocities are 
given in atomic units, where $\hbar = 1, m_e = 1, e = 1$ and the unit of atomic velocity is $v_e = 
\alpha c \approx \frac{c}{137} \approx 2.1882661 \cdot 10^{8}$ $cm \cdot sec^{-1}$. Here and 
everywhere below $c$ is the speed of light in vacuum and $\alpha = \frac{e^2}{\hbar c}$ is the 
dimensionless fine structure constant. This `atomic velocity' $v_e$ is the velocity of the 
$1s-$electron in the hydrogen atom with the infinitely heavy nucleus ${}^{\infty}$H. It is clear 
that in atomic units $v_e = 1$.

Let us evaluate the probabilities of formation of the negatively charged tritium and protium 
ions. In other words, we want to determine the probabilities of formation of the two-electron 
${}^3$H$^{-}$ and ${}^1$H$^{-}$ ions during the nuclear reaction, Eq.(\ref{en1}), in the 
two-electron ${}^3$He atom. According to the theory of sudden approximations and in respect 
with Eq.(\ref{eq3}) such a probability of formation of the tritium ion (${}^3$H$^{-}$ or 
T$^{-}$) is written in the form $P_{if} = \mid A_{if} \mid^2$, where $A_{if}$ is the probability 
amplitude which is written in the form
\begin{eqnarray}
 A_{if} = \langle \Phi_{{{\rm T}^{-}}}({\bf r}_1, {\bf r}_2) \mid \Psi^{\prime}_{{\rm He}}({\bf r}_1, 
 {\bf r}_2) \rangle = \langle \Phi_{{\rm T}^{-}}({\bf r}_1, {\bf r}_2) \mid 
 exp(\imath {\bf V}_t \cdot {\bf r}_1 + \imath {\bf V}_t \cdot {\bf r}_2) 
 \Psi_{{\rm He}}({\bf r}_1, {\bf r}_2) \rangle \label{eq45}
\end{eqnarray}
where $V_t$ is the speed of the tritium nucleus after the reaction, Eq.(\ref{en1}). In other words,
the probability amplitude is the overlap integral between the tritium ion and helium-3 wave 
functions, but the wave function of the helium-3 atom must be taken in the moving reference frames.  

In the incident and final wave functions we can separate three internal variables $r_{32}, r_{31}, 
r_{12}$ (or relative coordinates $r_{ij} = \mid {\bf r}_{i} - {\bf r}_{j} \mid = \mid {\bf r}_{ij} 
\mid$) from other six (3 + 3) variables which correspond to the translational and rotational degrees 
of freedom of the whole three-body system. Here and everywhere below the notations 1 and 2 mean the 
electrons, while the notation/index 3 stands for the central (heavy) nucleus. In atoms with one 
heavy nucleus the internal coordinates coincide with the interparticle, or relative coordinates. 
In general, the expression for the probability amplitude $A_{if}$, Eq.(\ref{eq45}), is reduced to 
the following form  
\begin{eqnarray}
 A_{if} &=& \int \int \int {\cal Y}^{{{\rm T}^{-}}}_{LM}({\bf r}_{31},{\bf r}_{32}) 
 \Phi_{{{\rm T}^{-}}}(r_{32}, r_{31}, r_{21}) 
 exp(\imath {\bf V}_t \cdot {\bf r}_{32} + \imath {\bf V}_t \cdot {\bf r}_{31}) \times \nonumber \\
 & & {\cal Y}^{{\rm He}}_{LM}({\bf r}_{31},{\bf r}_{32}) \Psi_{{\rm He}}(r_{32}, r_{31}, r_{21}) 
 r_{32} r_{31} r_{21} dr_{32} dr_{31} dr_{21} \label{eq5}
\end{eqnarray}
for the tritium ${}^3$H$^{-}$ (or T$^{-}$) ion. The notations ${\cal Y}^{{{\rm T}^-}}_{LM}({\bf 
r}_{31},{\bf r}_{32})$ and ${\cal Y}^{{\rm He}}_{LM}({\bf r}_{31},{\bf r}_{32})$ used in this equation
designate the corresponding bi-polar harmonics \cite{Var}, \cite{FrEf}. They are taking care about 
non-zero angular momenta of the incident and final atomic species. It should be mentioned here that the 
negatively charged hydrogen ion H$^{-}$ has only one bound $1^1S(L = 0)-$state. Also, in this study we 
restrict ourselves to the case when the incident ${}^3$He atom was in its ground $1^1S(L = 0)-$state. 
In this case all bipolar harmonics in Eq.(\ref{eq5}) equal unity and the probability amplitude, 
Eq.(\ref{eq5}), takes the form
\begin{eqnarray}
 & & A_{if} = \int \int \int \Phi_{{\rm T}^{-}}(r_{32}, r_{31}, r_{21}) j_0(V_t \cdot r_{32}) j_0(V_t 
 \cdot r_{31}) \Psi_{{\rm He}}(r_{32}, r_{31}, r_{21}) r_{32} r_{31} r_{21} dr_{32} dr_{31} 
 dr_{21} \nonumber \\ 
  &=& \frac{1}{V^2_t} \int \int \int \Phi_{{\rm T}^{-}}(r_{32}, r_{31}, r_{21}) sin(V_t \cdot r_{32}) 
 sin(V_t \cdot r_{31}) \Psi_{{\rm He}}(r_{32}, r_{31}, r_{21}) r_{12} dr_{32} dr_{31} dr_{21} 
 \label{eq6}
\end{eqnarray}
where $V_t$ is the speed of the tritium nucleus after the nuclear reaction in the ${}^3$He atom. 

The wave functions of the ground $1^1S(L = 0)-$states in the two-electron H$^{-}$ ion and He atom 
are usually approximated with the use of highly accurate variatonal expansion written in the 
relative/perimetric coordinates $r_{32}, r_{31}$ and $r_{21}$ or $u_1, u_2, u_3$ (more details can
be found, e.g., \cite{Fro01}). The most advanced of such expansions is the exponential variational 
expansion in the relative coordinates. It takes the following form (for the bound $S(L = 0)-$states 
in  the two-electron systems):
\begin{eqnarray}
 \psi(r_{32}, r_{31}, r_{21}) = \frac{1}{\sqrt{2}} [1 + (-1)^{\kappa} \hat{P}_{12}] \sum^{N}_{i=1} C_i 
 exp(-\alpha_i r_{32} - \beta_i r_{31} - \gamma_i r_{21}) \chi(1,2) \label{eq7}
\end{eqnarray} 
where $C_i$ are the linear variational coefficients, $\hat{P}_{12}$ is the permutation of the two 
identical particles (electrons 1 and 2) and $N$ is the total number of terms in the trial function 
$\psi(r_{32}, r_{31}, r_{21})$ which is an accurate approximation of the actual wave function 
$\Psi(r_{32}, r_{31}, r_{21})$. In Eq.(\ref{eq7}) the notation $\chi(1,2)$ stands for the 
two-electron spin function. For the singlet states one needs to chose $\chi(1,2) = \frac{1}{\sqrt{2}} 
(\alpha \beta - \beta \alpha)$ and $\kappa = 0$ in Eq.(\ref{eq7}). The total energies obtained for the 
ground $1^1S-$states of the H$^{-}$ ion and He atom with the use of Eq.(\ref{eq7}) can be found in 
Table I. The wave functions, Eq.(\ref{eq7}), are used in calculations of the probability to form the 
bound tritium ion ${}^3$H$^{-}$.
 
If the $\Phi_{{\rm T}^{-}}(r_{32}, r_{31}, r_{21})$ and $\Psi_{{\rm He}}(r_{32}, r_{31}, r_{21})$ in 
Eq.(\ref{eq6}) are represented in the form of Eq.(\ref{eq7}), then the probability amplitude $A_{if}$ 
is written as the double sum of the following three-particle integrals 
\begin{eqnarray}
 & & B^{(00)}_{0;0;1}(a, b, c; V_t) = \frac{1}{V^{2}_{t}} \int \int \int exp(-a r_{32} - b r_{31} - 
 c r_{21}) sin(V_t \cdot r_{32}) \times \nonumber \\
 & & sin(V_t \cdot r_{31}) r_{12} dr_{32} dr_{31} dr_{21} \label{eq8}
\end{eqnarray}
where $a = \alpha_i({{\rm T}^{-}}) + \alpha_j({{\rm He}}), b = \beta_i({{\rm T}^{-}}) + 
\beta_j({{\rm He}})$ and $c = \gamma_i({{\rm T}^{-}}) + \gamma_j({{\rm He}})$. Theory of these integrals 
was developed in \cite{Fro2013}. In particular, it was shown in \cite{Fro2013} that such an integral is 
reduced to the following double sum (here we apply the Cauchy formula) 
\begin{eqnarray}
 B^{(00)}_{0;0;1}(a, b, c; V) = \sum^{\infty}_{\kappa=0} \frac{(-1)^{\kappa} V^{2 
 \kappa}}{(2 \kappa + 2)!} \sum^{\kappa}_{\mu=0} C^{2 \mu + 1}_{2 \kappa + 2} 
 \Gamma_{2 \mu + 1;2 \kappa - 2 \mu + 1;1}(a, b, c) \label{eq46}
\end{eqnarray}
where $C^{k}_{n}$ is the binomial coefficient, i.e. the number of combinations from $n$ by $k$ ($n 
\ge k$), and $\Gamma_{k;l;n}(a,b,c)$ is the basic three-particle integral defined in \cite{Fro2013}. 
This formula allows one to determine the probability to form the bound T$^{-}$ (or ${}^3$H$^{-}$) 
and ${}^1$H$^{-}$ ions during the nuclear reaction, Eq.(\ref{en1}), in the two-electron ${}^3$He atom. 
For instance, by using the approximate one-term wave functions for the ground state in the helium 
atom and hydrogen ion given in Table II of Ref.\cite{Fro1} we have found that the probability to form
the bound T$^{-}$ (or ${}^3$H$^{-}$) ion in the reaction Eq.(\ref{en1}) is $\approx$ 0.77048798 \%
(probability amplitude is $\approx$ 0.87777445$\cdot 10^{-1}$). Such a large probability of the T$^{-}$ 
ion formation means that these ions formed in the reaction, Eq.(\ref{en1}), can be dected in modern 
experiments. Analogous probability for the negatively charged protium ion is only $\approx$ 
2.391074$\cdot 10^{-5}$ \% (probability amplitude is $\approx$ 0.48898613$\cdot 10^{-3}$), i.e. it is 
significantly smaller. This illustrates a very strong dependence of the final state probabilities upon 
the velocity $V$ of the final atomic fragment, if $V \ge 1$ \cite{Mig1}, \cite{Fro1}.

It is interesting to note that we can also use the formula, Eq.(\ref{eq46}), in the case when $V = 0$. This
case corresponds to the $\beta^{-}$ decay of the ${}^3$H$^{-}$ ion into the two-electron ${}^3$He atom.
The corresponding probability obtained with our one-term wave functions is $\approx$ 23.893045 \% 
(probability amplitude is $\approx$ 0.4880502483). These amplitude and final probability are very close 
to our earlier prediction made in 1998 \cite{Fro981}. Analogous calculations with the use of five-term 
variational wave functions (with the carefuly optimized non-linear parameters) for the H$^{-}$ ion and 
He atom gives the following probabilities: 21.075287 \% ($\beta^{-}$ decay), 0.581089 \% (${}^3$H$^{-}$
ion formation) and 1.842681$\cdot 10^{-5}$ \% (${}^1$H$^{-}$ ion formation). The corresponding variational 
three-term energies for the ${}^{\infty}$H$^{-}$ ion and ${}^{\infty}$He atom are -0.5277402583285 $a.u.$ 
and -2.903691563543 $a.u.$, respectively (compare with the `exact' energies from Table I).

Note that our variational few-term wave functions constructed for the H$^{-}$ ion and He atom by uisng 
with carefully optimized non-linear parameters are the best functions in their class. Nevertheless, it
is very interesting to check our predictions by applying variational wave functions for the H$^{-}$ ion 
and He atom with significantly larger number of terms, e.g., the trial wave functions from Table I. 
However, right now such calculations cannot be performed, since there is an additional problem here 
related with the use of the non-orthogonal basis sets in calculations of the overlap integrals which 
include the two different wave functions. This problem was never discovered in earlier studies where 
different expectation values were always computed for the same systems. In such cases the bound state 
wave functions are exactly the same (i.e. identical) for the `incident' and `final' state. For instance,
by using our trial wave functions written in the non-orthogonal basis we can determine the probabilities
of `non-excitation' during the two following processes
\begin{eqnarray}
 &&{}^3{\rm He}(1^1S; V = 0) + n =  {}^3{\rm He}(1^1S; V) + n^{\prime} \label{eq9} \\
 &&{}^3{\rm H}^{-}(1^1S; V = 0) + n =  {}^3{\rm H}^{-}(1^1S; V) + n^{\prime} \label{eq91}
\end{eqnarray} 
Briefly, the atomic nuclei of the ${}^3$He atom and ${}^3$H$^{-}$ ion are suddenly accelerated to the 
final speed $V$ by fast neutrons. Such processes were studied in detail in \cite{TalFro06}. The question 
is to evaluate the probabilities of the incident atoms to stay in the same $1^1S$-states (ground states) 
and keep two bound electrons. Such probabilities have been determined with the method described above and 
variational wave functions which contain up to 400 terms. The results for the ${}^3$He atom and ${}^3$H$^{-}$ 
ions can be found in Table II for $V$ = 0.0, 1.0, 1.59632, 2.0, 3.5 and 4.78797 (all these velocities are 
exressed in atomic units ($a.u.$)). The computed probabilities are numerically stable and they converge when
the total number of basis functions $N$ in Eq.(\ref{eq7}) increases. Note again that in these cases the 
incident and final wave functions are identical to each other.

This situation changes drastically, if we consider the overlap integrals of two different wave functions 
written in the non-orthogonal basis sets. We do not know the correct order of non-orthogonal basis vectors 
in both wave functons and this leads to some serious problems in calculations. For instance, let us change 
the order of basis vectors in the He-wave function, Eq.(\ref{eq7}). It does not change the norm of this wave 
function, but numerical value of the overlap integral with the tritium ion wave function will be changed 
substantially, since the transformation which connects these two non-orhtogonal basis sets is not unitary 
(or orthogonal). Briefly, this means that we cannot predict the correct order of basis functions in the two 
approximate wave functions. But only such an order is appropriate for correct calculations of the final 
state probabilities. Some simple method recomended for solution of this problem, e.g., the use of natural 
expansions of the wave functions and/or addition orthogonalization of these functions, either lead to a 
substantial loss of overall numerical accuracy, or to other problems. At this moment we are trying to 
solve this intersting problem and develop the new, reliable method for accurate computations of the final 
state probabilities.    

For the first time we have developed the closed and transparent procedure which allows one to calculate 
the probabilities of formation of few-electron atomic species during nuclear reactions in the incident 
atom(s). To evaluate such probabilities in earlier studies we have to apply approximate procedures with
model (one-electron) wave functions. By using the results of this study we evaluated the probability of 
formation of the tritium (${}^{3}$H$^{-}$) ion in the nuclear reaction, Eq.(\ref{en1}), $\approx$ 0.8 - 1\% 
(not 8 \% as it was predicted earlier). The approach developed here can be applied to more complicated 
cases, e.g., to the reaction of the ${}^{10}$B nuclei with slow neutrons in the five-electron boron atom 
(this reaction is used in the boron neutron capture therapy to treat cancer).

\newpage
\begin{table}[tbp]
   \caption{The total energies $E$ of the ground $1^1S(L = 0)-$states in the negatively charged 
            hydrogen ion ${}^{\infty}$H$^{-}$ and ${}^{\infty}$He atom (in atomic units). $K$ 
            is the total number of basis functions used.}
     \begin{center}
%     \scalebox{0.72}{%
     \begin{tabular}{| c || c || c |}
      \hline\hline
 $K$ & $E$(${}^{\infty}$H$^{-}$) & $E$(${}^{\infty}$He) \\
     \hline
 3500 & -0.527751 016544 377196 590213 & -2.903724 377034 119598 030965 \\
 
 3700 & -0.527751 016544 377196 590333 & -2.903724 377034 119598 030983 \\

 3840 & -0.527751 016544 377196 590389 & -2.903724 377034 119598 030995 \\

 4000 & -0.527751 016544 377196 590446 & -2.903724 377034 119598 031033 \\
    \hline\hline
  \end{tabular}
  \end{center}
  \end{table}
\begin{table}[tbp]
   \caption{The probabilities of non-excitation $p_f$ (in \%) for the processes, Eq.(\ref{eq9}) and 
            Eq.(\ref{eq91}), with the ${}^{3}$He atom and ${}^{3}$H$^{-}$ ion. $V$ is the final 
            velocity (in $a.u.$) of the central atomic nucleus.}
     \begin{center}
%     \scalebox{0.95}{%
     \begin{tabular}{| l || l | l | l | l | l | l |}
      \hline\hline
 $V$             & 0.0 & 1.0 & 1.59632 & 2.0 & 3.5 & 4.78797 \\
     \hline
 ${}^{3}$He      & 100.0 & 46.90946 & 16.53096 & 6.95123 & 0.17532 & 7.48683$\cdot 10^{-3}$ \\
 
 ${}^{3}$H$^{-}$ & 100.0 & 1.217223 & 2.27539$\cdot 10^{-2}$ & 2.61097$\cdot 10^{-3}$ & 4.2066$\cdot 10^{-6}$ &
                              6.0222$\cdot 10^{-9}$ \\
    \hline\hline
  \end{tabular}
  \end{center}
  \end{table}

\begin{thebibliography}{10}

\bibitem{Ehre} P. Ehrenfest, Zs. Phys. \textbf{45}, 455 (1927).

\bibitem{LLQ} L.D. Landau and E.M. Lifshitz, {\it Quantum Mechanics. Non-Relativistic 
Theory}, (3rd. edn., Pergamonn Press, New York (1977)).

\bibitem{Kaemp} F.A. Kaempffer, {\it Concepts in Quantum Mechanics}, (Academic Press, 
New York (1965)), Chps. 9 and 12.

\bibitem{Mig1} A.B. Migdal, J. Phys. (USSR) \textbf{4}, 449 (1941).

\bibitem{Mig2}A.B. Migdal and V. Krainov, \textit{Approximation Methods in
Quantum Mechanics}, (W.A. Benjamin Inc., New York (1969)).

\bibitem{Kik} \textit{Handbook. Tables of Physical Quantaties} (Ed. I.K.
Kikoin, Atomizdat, Moscow (1974)), (in Russian).

\bibitem{Gon} G.A. Goncharov, Phys. Today {\bf 49}, 60 (1996).

\bibitem{Fro98} A.M. Frolov, Plasma Phys. Contr. Fusion \textbf{40}, 1417
(1998).

\bibitem{Fro1} A.M. Frolov and D.M. Wardlaw, Phys. Rev. A \textbf{79},
032703 (2009).

\bibitem{Fro2} A.M. Frolov and D.M. Wardlaw, J. Phys. B \textbf{44},
105005 (2011).

\bibitem{WW} A.M. Weinberg and E.P. Wigner, \textit{The Physical Theory of
Neutron Chain Reactors} (University of Chicago Press, Chicago IL (1959)).

\bibitem{Var} D.A. Varshalovich, A.N. Moskalev and V.K. Khersonskii, {\it 
Angular Momentum in Quantum Mechanics. Non-Relativistic Theory}, (3rd. edn. 
Pergamonn Press, New York (1977)).

\bibitem{FrEf} A.M. Frolov and V.D. Efros, Pis'ma Zh. Eksp. Teor. Fiz. {\bf 39}, 449 
(1984)[Sov. Phys. JETP Lett., {\bf 39}, 544 (1984)].

\bibitem{Fro01} A.M. Frolov, Phys. Rev. E {\bf 64}, 036704 (2001).

\bibitem{Fro2013} A.M. Frolov, {\it Three-particle integrals with the Bessel 
functions}, arXiv: 1211.4536 (2012).

\bibitem{Fro981} A.M. Frolov, Phys. Rev. A {\bf 58}, 4479 (1998).

\bibitem{TalFro06} J.D. Talman and A.M. Frolov, Phys. Rev. A {\bf 73}, 032722 (2006).

\end{thebibliography}
\end{document}